\documentclass[runningheads]{cl2emult}

\usepackage{graphicx} 

\begin{document}
\title*{Neutron Star Mass Determinations}
\author{M.~H.~van Kerkwijk}
\institute{Utrecht University, P.~O.~Box 80000, 
           3508~TA Utrecht, The Netherlands} 

\maketitle              

\begin{abstract}
I review attempts made to determine the properties of neutron stars.
The focus is on the maximum mass that a neutron star can have, or,
conversely, the minimum mass required for the formation of a black
hole.  There appears to be only one neutron star for which there is
strong evidence that its mass is above the canonical $1.4\,M_\odot$,
viz., Vela~X-1, for which a mass close to $1.9\,M_\odot$ is found.
Prospects for progress appear brightest for studies of systems in
which the neutron star should have accreted substantial amounts of
matter.
\end{abstract}

\section{The Mimimum Mass Required to Form a Black Hole}
For the study of black holes, the relevance of neutron stars is mostly
that below a certain maximum mass, degeneracy pressure due to nucleons
is sufficient to prevent an object from becoming a black hole.
Unfortunately, the equation of state of matter at densities above
nuclear-matter density is rather uncertain and theoretical estimates
of the minimum mass required to form a black hole range from just over
$1.4\,M_\odot$ to about  $2.5\,M_\odot$ \cite{d88,s00}.

Constraints on the equation of state can be obtained from a variety of
observed properties of neutron stars.  Among the more direct
measurements that have been used or proposed are: (i) the maximum mass
inferred from dynamical measurements in binaries; (ii) the minimum
spin period among milli-second pulsars; (iii) identification of kHz
quasi-periodic oscillations with the orbital frequency in the last
stable orbit; (iv) identification of quasi-periodic oscillations with
the Lens-Thirring precession period; (v) influence of gravitational
light bending on X-ray light curves in radio pulsars; (vi)
model-atmosphere fitting of X-ray lightcurves during X-ray bursts
(resulting from to thermonuclear run-aways on the neutron-star
surface); (vii) gravitational redshift from $\gamma$-ray spallation
lines in accreting systems; (viii) gravitational redshift and surface
gravity from model-atmosphere analysis of spectra of isolated neutron
stars.  Less direct measurements include: (ix) matching observed
neutron-star temperatures and inferred ages to cooling curves; (x)
neutrino light curves in supernova explosions; (xi) pulsar glitches;
(xii) moment of inertia from accretion torques in combination with
knowledge of the magnetic field strength from cyclotron lines; (xiii)
comparing X-ray fluxes between states in which a neutron star is
accreting and in which matter is stopped at the magnetosphere and
``propelled'' away.  For references and somewhat more detail, see
\cite{vp98}.  The strongest constraints on the equation of state are
still set by dynamical mass measurements, so I will restrict myself to
those below.

\section{Dynamical measurements}
Most mass determinations have come from radio timing studies of
pulsars; see \cite{tc98} for an excellent review.  The most accurate
ones are for pulsars that are in eccentric, short-period orbits with
other neutron stars, in which several non-Keplerian effects on the
orbit can be observed: the advance of periastron, the combined effect
of variations in the second-order Doppler shift and gravitational
redshift, the shape and amplitude of the Shapiro delay curve shown by
the pulse arrival times as the pulsar passes behind its companion, and
the decay of the orbit due to the emission of gravitational waves.
The most famous of the double neutron-star binaries is the
Hulse-Taylor pulsar, PSR B1913+16, for which recent measurements give
$M_{\rm{}PSR}=1.4411\pm0.0007\,M_\odot$ and
$M_{\rm{}comp}=1.3874\pm0.0007\,M_\odot$ \cite{tw89,t92}.  Almost as
accurate masses have been inferred for PSR B1534+12, for which the
pulsar and its companion are found to have very similar mass: for
both, $M=1.339\pm0.003\,M_\odot$ \cite{sac+98}.

Neutron-star masses can also be determined for some binaries
containing an accreting X-ray pulsar, from the amplitudes of the X-ray
pulse delay and optical radial-velocity curves in combination with
constraints on the inclination (the latter usually from the duration
of the X-ray eclipse, if present).  This method has been applied to
about half a dozen systems \cite{jr84,n89,vkvpz95}, but the masses are
generally not very precise.

So far, for all but one of the neutron stars, the masses are
consistent with being in a surprisingly narrow range, which can be
approximated with a Gaussian distribution with a standard deviation of
only $0.04\,M_\odot$ \cite{tc98}.  The mean of the distribution is
$1.35\,M_\odot$, close to the ``canonical'' value of $1.4\,M_\odot$.

The one exception is the X-ray pulsar Vela~X-1, which is in a 9-day
orbit with the B0.5\,Ib supergiant HD~77581.  For this system, a
rather higher mass of around $1.8\,M_\odot$ has
consistently\footnote{One study based on IUE spectra of HD~77581
appeared to find a lower neutron-star mass \cite{slrw97}.  However,
this was found to be due to a bug in the cross-correlation software
used \cite{bkvk+00}} been found ever since the first detailed study in
the late seventies \cite{vpzt+77,vkvpz+95}.  A problem with this
system is that the measured radial velocities show strong deviations
from a pure Keplerian radial-velocity curve, which are correlated
within one night, but not from one night to another.  A possible cause
could be that the varying tidal force exerted by the neutron star in
its eccentric orbit excites high-order pulsation modes in the optical
star which interfere constructively for short time intervals.

\begin{figure}
\centering
\includegraphics[width=0.8\textwidth]{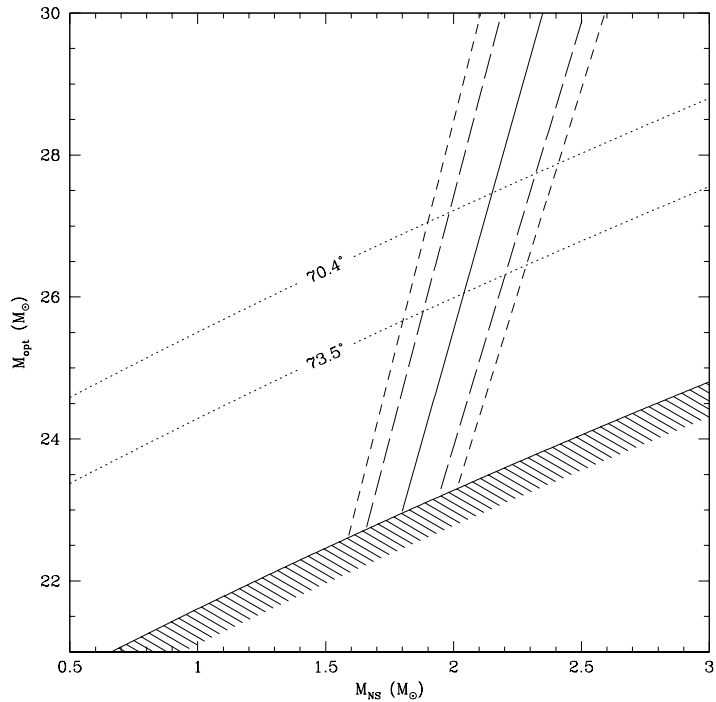}
\caption[]{Constraints on the mass of Vela~X-1 and its supergiant
companion HD~77581 \cite{bkvk+00,bcc+97}.  The constraint on the mass ratio from the X-ray
pulse delay and optical radial-velocity curves is indicated by the
solid line.  The long and short-dashed lines next to it indicate the
95\% and 99\% confidence limits, respectively.  The lines stop at the
region excluded by the pulse-timing mass function (to go below it
would require $\sin{}i>1$).  The 95\% and 99\% confidence lower limits
on the inclination derived from the duration of the X-ray eclipse are
indicated by the two dotted lines}
\label{fig:vela}
\end{figure}

We were granted time at ESO to improve the mass determination of this
possibly very massive neutron star from 200 new spectra, taken in as
many nights.  These cover more than 20 orbits, and make it possible to
average out the velocity excursions and to constrain possible
systematic effects with orbital phase.  In combination with
measurements from our old photographic plates, earlier CCD
spectroscopy, and high resolution IUE spectra, we derived a 95\%
confidence constraint on the mass of the neutron star of
$M_{\rm{}NS}=1.87^{+0.23}_{-0.17}$ \cite{bkvk+00}.  Our constraints
are illustrated graphically in Fig.~\ref{fig:vela}.  One sees that
even at 99\% confidence, $M_{\rm{}NS}>1.6\,M_\odot$.  It should be
noted, however, that from the data it appears that while the
excursions in radial velocity are mostly random, there is also
a component that is systematic, locked to orbital phase.  Since we do
not understand these effects, it may be that our mass estimate is
biased.  In our trials with excluding the worst-affected phase ranges,
however, we consistently found that the fitted mass became even higher
\cite{vkvpz+95,bkvk+00}. 

\section{Trying for Bias}
The narrow range in masses inferred for most neutron stars might be
seen as evidence for a relatively low maximum neutron-star mass.  It
could also be, however, that it reflects the formation mechanism.
Indeed, from models, it appears that supernova explosions result in
neutron stars with masses preferentially in two narrow peaks, one
around $1.3\,M_\odot$ and one around $1.65\,M_\odot$ \cite{tww96}.  It
is tempting to associate the latter with Vela~X-1, but for honesty it
should be noted that from the same calculations it is expected that
only lighter neutron stars can be formed by stars which lose their
envelope during their evolution, as would likely have happened for the
progenitor of Vela~X-1.

If the narrow range indeed reflects the formation process, it seems
worthwile to focus especially on systems in which the neutron star is
expected to have accreted a lot of matter since it was formed.  Such
systems are the low-mass X-ray binaries and their descendents, the
pulsars with low-mass white dwarf companions.  In the latter systems,
the white dwarfs typically have masses of $0.3\,M_\odot$.  However, in
order for mass transfer to have happened, these stars need to have
evolved to at least the end of the main sequence life.  For this to
happen in a Hubble time, their masses need to have been at least
$0.8\,M_\odot$ initially.  Since the mass transfer in these systems is
thought to be stable, the neutron star should thus have accreted more
than $0.6\,M_\odot$, i.e., have become substantially more massive than
it was initially.

\begin{figure}
\centering
\includegraphics[width=0.8\textwidth]{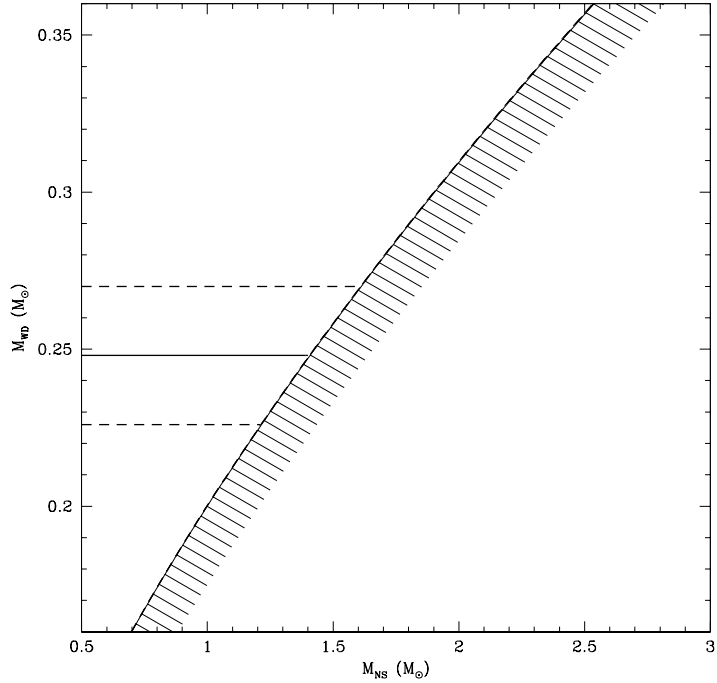}
\caption[]{Constraints on the mass of PSR~B1855+09 and its white-dwarf
companion \cite{ktr94,tc98}.  The horizontal line is at the best fit
white-dwarf mass derived from the amplitude of the Shapiro delay
curve, and the short-dashed lines reflect the 95\% confidence
uncertainties on that measurement.  The lines stop at the limit
derived from the pulse timing mass function, at $i=90^\circ$; to be to
the right or below it would require $\sin{}i>1$.  The barely visible
short-dashed curve just left of it reflects the 95\% confidence lower
limit on the inclination set by the shape of the observed Shapiro
delay curve (the upper limit would not be visible in this graph)}
\label{fig:psrb1855}
\end{figure}

From radio timing measurements, it is generally more difficult to
measure masses for these systems, because the orbits are circular and
relatively wide.  However, for one system, PSR B1855+09, the orbit is
very close to edge on, and it has been possible to derive constraints
from the Shapiro delay \cite{ktr94,tc98}.  These constraints are
indicated in Fig.~\ref{fig:psrb1855}.

\begin{figure}
\centering
\includegraphics[width=0.8\textwidth]{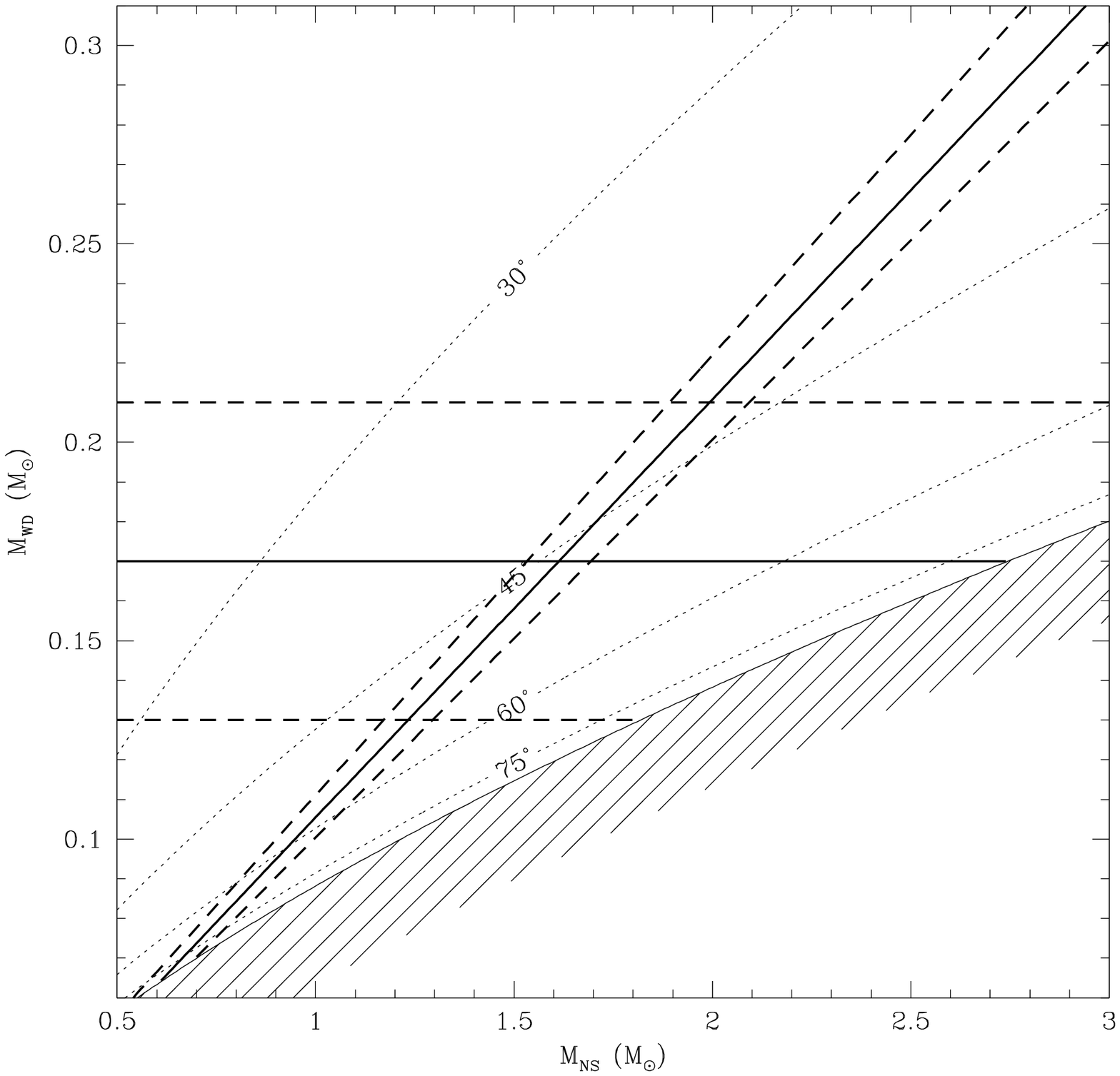}
\caption[]{Constraints on the mass of PSR~J1012+5307 and its
white-dwarf companion \cite{vkbk96,cgk98}.  The horizontal solid line
reflect the white-dwarf mass inferred from the surface gravity
measured from the optical spectrum of the white dwarf.  The slanted
solid line is the constraint set by the mass ratio inferred from the
optical radial-velocity and radio pulse-delay curves.  For both
curves, the 95\% confidence regions are indicated by the associated
short-dashed lines.  All lines stop at the limit derived from the
pulse timing mass function, at $i=90^\circ$; to be to the right or
below it would require $\sin{}i>1$.  Dotted lines indicate the
relations expected for some other values of the inclination}
\label{fig:psrj1012}
\end{figure}

Another way of determining the masses in these systems uses optical
spectroscopy of the white-dwarf companion.  From a model-atmosphere
fit to the spectrum, one can determine the effective temperature and
surface gravity.  From the latter, the white-dwarf mass follows,
assuming a theoretical mass-radius relation.  If one can also measure
the radial-velocity orbit, and determine the mass ratio (in
combination with the pulse-delay orbit), then one has a constraint on
the neutron-star mass.  So far, the only pulsar for which this has
been possible is PSR~J1012+5307 \cite{vkbk96,cgk98}, whose companion
is particularly bright ($V\simeq20$).  The present constraints are
shown in Fig.~\ref{fig:psrj1012}.

One sees that for both systems, the mass determinations are at present
not precise enough to provide meaningful additional constraints on the
maximum mass a neutron star can have.  The situation unfortunately is
no better for determinations in low-mass X-ray binaries.  A problem
for those is that the X-ray sources generally do not pulse, and hence
all information has to be gleaned from observations of the companions.
So far, a meaningful determination has been possible only for Cyg~X-2
\cite{cck97,ok99}.  A relatively high mass is found, but again the
uncertainties are rather large.

Despite the above, prospects for progress appear brightest for further
studies of systems in which the neutron star should have accreted a
substantial amount of matter.  Some other white-dwarf companions are
bright enough, and more may be found in the on-going pulsar searches.


\end{document}